\documentclass[5p,times]{elsarticle}

\usepackage[utf8]{inputenc}
\usepackage{hyperref}
\usepackage{color}
\usepackage{comment}
\usepackage{svn-multi}
\usepackage{amsmath}
\usepackage{calc}
\svnidlong
{$HeadURL: https://solps-mdsplus.aug.ipp.mpg.de/repos/ASCOT/branches/papers/SOFT2014-riplos2comsol/akaslompolo_soft2014_4pp.tex $}
{$LastChangedDate: 2015-05-21 10:58:08 +0300 (Thu, 21 May 2015) $}
{$LastChangedRevision: 7405 $}
{$LastChangedBy: sjjamsa $}
\journal{Fusion Engineering and Design, \url{http://dx.doi.org/10.1016/j.fusengdes.2015.05.038}}

\begin{document}

\begin{frontmatter}

\title{Calculating the 3D magnetic field of ITER for European TBM studies}

\author[aalto]{Simppa \"Ak\"aslompolo\corref{mycorrespondingauthor}}
\ead{simppa.akaslompolo@alumni.aalto.fi}
\author[aalto]{Otto Asunta}
\author[aalto]{Thijs Bergmans}
\author[f4e]{Mario Gagliardi}
\author[f4e]{Jose Galabert}
\author[aalto]{Eero Hirvijoki}
\author[aalto]{Taina Kurki-Suonio}
\author[aalto]{Seppo Sipil\"a}
\author[aalto]{Antti Snicker}
\author[aalto]{Konsta S\"arkim\"aki}
\cortext[mycorrespondingauthor]{Corresponding author}

\address[aalto]{Department of Applied Physics, Aalto University, FI-00076 AALTO, FINLAND}
\address[f4e]{Fusion for Energy, Barcelona, Spain}

\begin{abstract}
The magnetic perturbation due to the ferromagnetic test blanket
modules (TBMs) may deteriorate fast ion confinement in ITER. This
effect must be quantified by numerical studies in 3D. We have
implemented a combined finite element method (FEM) -- Biot-Savart law
integrator method (BSLIM) to calculate the ITER 3D magnetic field and
vector potential in detail. Unavoidable geometry
simplifications changed the mass of the TBMs and ferritic inserts
(FIs) up to 26\%.  This has been compensated for by modifying the
nonlinear ferromagnetic material properties accordingly. Despite the
simplifications, the computation geometry and the calculated fields
are highly detailed. The combination of careful FEM mesh design and
using BSLIM enables the use of the fields unsmoothed for particle
orbit-following simulations. The magnetic field was found to agree
with earlier calculations and revealed finer details. The vector
potential is intended to serve as input for plasma shielding
calculations.

\end{abstract}


\end{frontmatter}


\section{Introduction}
\noindent The goal of the fusion reactor ITER is to demonstrate the technological and scientific feasibility of fusion energy. The following reactor, DEMO, has a mission to demonstrate the large-scale production of electrical power and tritium fuel self-suf\-ficien\-cy. One of the tasks of ITER is to be a testbed for DEMO components. ITER Test Blanket Modules (TBMs) will test the technology of tritium breeding modules for DEMO. Three of the eighteen ITER equatorial ports are reserved for these modules. The material chosen for DEMO is ferritic steel \cite{Boccaccini2011478}. Therefore, the ITER TBMs will be made of ferromagnetic material that will get magnetized by the tokamak magnetic fields. The resulting local perturbation in the magnetic fields can deteriorate the confinement of the plasma. Especially, the weakly collisional fast ions may find a local ``hole'' in the magnetic bottle and thus cause a hot spot on the first wall \cite{KramerNF}.

The TBM designs are checked for such threats by performing simulations of fast ions \cite{ascot_wall2009,shinohara2011}. The key ingredient for the simulations is a 3D magnetic field that includes the fields due to the magnetized components. \emph{This paper describes a method for calculating the field due to ferromagnetic ITER components. The used geometry describes the ferritic inserts and test blanket modules in detail.} We also calculate another useful quantity: the magnetic vector potential -- an important input for related plasma response calculations.

\paragraph{Input data} The key ``raw'' data we use to calculate the 3D field consists of the geometry and electric current in toroidal field (TF) and poloidal field (PF) coils (including the central solenoid), the plasma equilibrium, and the geometry and material parameters of the ferromagnetic components. Two sets of components are considered: the TBMs and the ferritic inserts (FI) that are in place to reduce the variation of the toroidal field caused by having only 18 discrete TF coils.

\paragraph{Methods} Our main tool is the commercial COMSOL Multiphysics finite element method (FEM) platform (ver. 4.4.0.195) and its AC/DC Module. For geometry simplifications we also used the SpaceClaim Engineer 3D direct modeler. In addition, MATLAB routines were used to prepare and deliver information to COMSOL as well as to build the COMSOL model.

No FEM calculation can match the precision of a direct Biot-Savart law integration for magnetic fields from known coil geometry.  To capitalise on this, we utilise a two-step COMSOL calculation: First we perform a \emph{magnetization calculation}. Then a follow-up \emph{permanent magnet} COMSOL calculation extracts the field due  to the magnetization from the results of the magnetization calculation. The permanent magnet results from COMSOL are finally superimposed on the Biot-Savart law integrated fields. These fields are produced with the recently extended BioSaw code \cite{Koskela12_ITER_ELM_coils_fast_ion_losses}.

In the magnetization calculation, the magnetic fields (mf) interface of the AC/DC module in COMSOL solves the vector potential $\mathbf{A}$ and the divergence control variable $\psi$ from
\begin{equation}
\left\{ \begin{array}{rcl}
\nabla\times \mathbf{H}(|\mathbf{B}|) &=& \mathbf{J_e}+\nabla\psi\\
 \nabla\times\mathbf{A} &=&\mathbf{B}\\
 \nabla\cdot\left(\mathbf{A}+\nabla\psi\right)&=&0
 \end{array}\right..
\end{equation}
Symbol $\mathbf{B}$ denotes magnetic flux density and $\mathbf{J_e}$ is the current density in coils or plasma. The nonlinear magnetic properties are described by the function $\mathbf{H}(|\mathbf{B}|)$. At the boundaries of the computational domain, the boundary condition is set to
$\mathbf{n}\times\mathbf{A}=0,$
where $\mathbf{n}$ is the boundary's normal vector. However, the boundaries are effectively several kilometres from the tokamak, as will be explained in the context of equation \ref{eq:inftyMapping}.

The permanent magnet model can be considered a reduced version of the above problem. We removed all coil and plasma currents and changed the constitutive relation inside the ferromagnetic components to $\mathbf{B}=\mu_0\left(\mathbf{H}+\mathbf{M}\right)$, where $\mathbf{M}$ is the magnetization vector. COMSOL extracts the magnetization from the solution of the first step and effectively turns the ferromagnetic components into permanent magnets. The divergence condition variable is used for the permanent magnet model only when the vector potential needs to be evaluated.

\section{Geometry}
\noindent In this section we describe each major component present in the FEM model. The description justifies and explains all simplifications made to the geometry before importing it to COMSOL.

The 18 ITER toroidal field coils are D-shaped superconducting coils enclosing the plasma. We reconstructed the 3D geometry by sweeping the coil cross section \cite{DDD11magnet} over the operating temperature spine curve \cite{TFWRADIALPLATE}. 
\label{sec:TFspineCurve}
However, the curve was not perfectly smooth and continuous after it was drawn according to the specifications. Therefore a 2\textsuperscript{nd} degree Bézier curve \cite{curves} was fitted to a dense sampling of the CAD curve. The numerically smooth final curve is in a format natively supported by COMSOL.

\begin{figure}
  \centering
  \raisebox{4.7cm}{(a)}\includegraphics[trim=7cm 1.2cm 7cm 0.2cm,clip=true,width=0.28\linewidth]{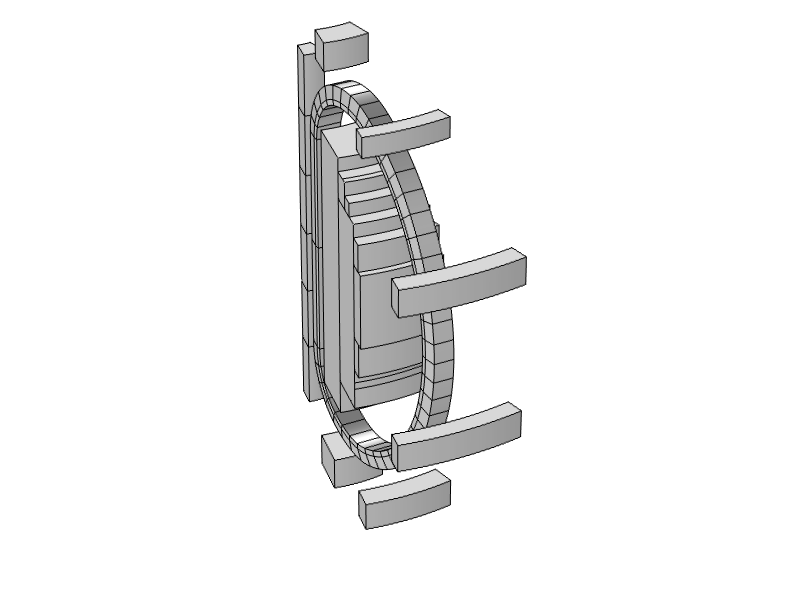}
  \raisebox{4.7cm}{(b)} \includegraphics[width=0.4\linewidth]{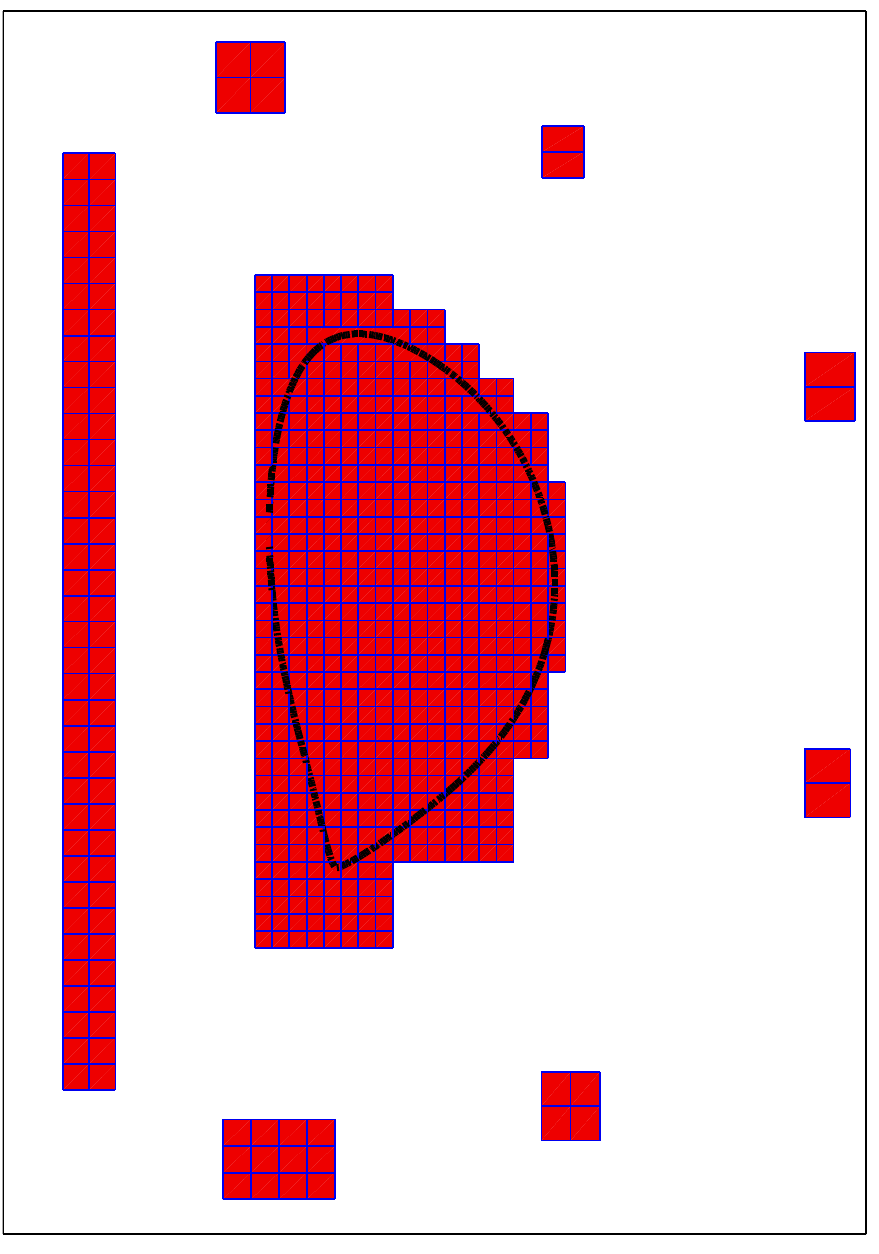}

\caption{
\label{fig:currents}
(a) A single 20 degree sector including all the current-carrying domains. The calculation was performed in the complete 360 degree geometry. (b) The cross-sectional mesh of the toroidal current-carrying domains. The thick round curve is the plasma separatrix. }
\end{figure}

The poloidal field (PF) coils, the central solenoid (CS) and the plasma are assumed to be toroidally symmetric.
 We received the geometry information in EQDISK format, a quasi-standard in fusion community. ITER has 6 CS coils and 6 PF coils. The only change to this input was the elimination of a 5 cm wide gap between the stacked CS coils by symmetrically stretching each coil vertically until they touched each other.
 Including this gap would have required small elements in the space between the coils without contributing significantly to the calculation. The current density was reduced to compensate for the increased cross-sectional area of the CS coils. The geometry is shown in Fig. \ref{fig:currents}.

The plasma cross-section was covered with a rectangular mesh (which was later swept toroidally) with the goal of extending the mesh up to the plasma-facing components (Fig. \ref{fig:currents}(b)). The mesh nodes were carefully fitted to coincide with the grid used in the EQDISK calculation, though the pitch was a multiple of the EQDISK grid pitch. This grid setting aligned the element edges to cylindrical coordinates. This was useful, as the current densities were now always parallel to an element. Furthermore, the calculated fields would later be exported in the same coordinate system.

\label{sec:FIgeom}

The FI geometry we received represented the ``configuration model'' of the components, including the special elements at the NBI ports. This is a simplified description primarily intended for checking that no two components in the vast ITER CAD model collide. Therefore small details such as bolt holes had already been removed. Nevertheless, the CAD data was still far too detailed for finite element analysis (FEA) as a part of a model encompassing the whole ITER torus.

\begin{figure}
  \centering
  \includegraphics[height=0.4\linewidth]{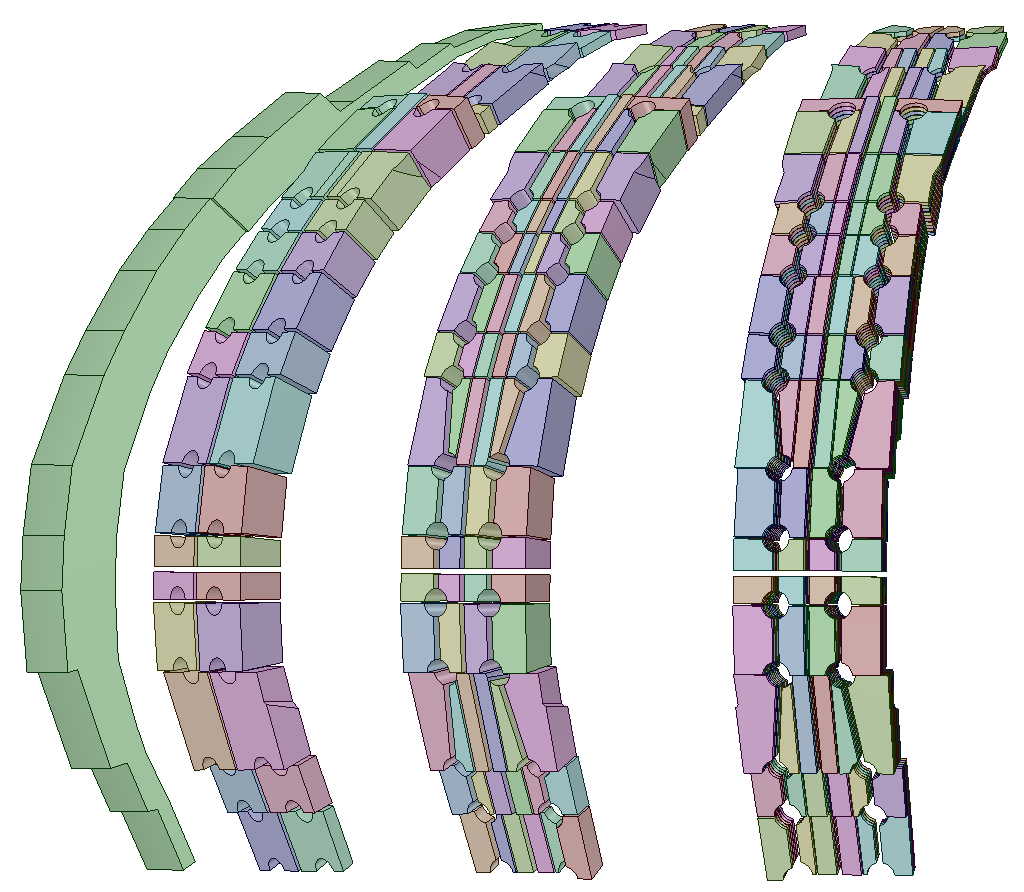}
  \includegraphics[height=0.4\linewidth]{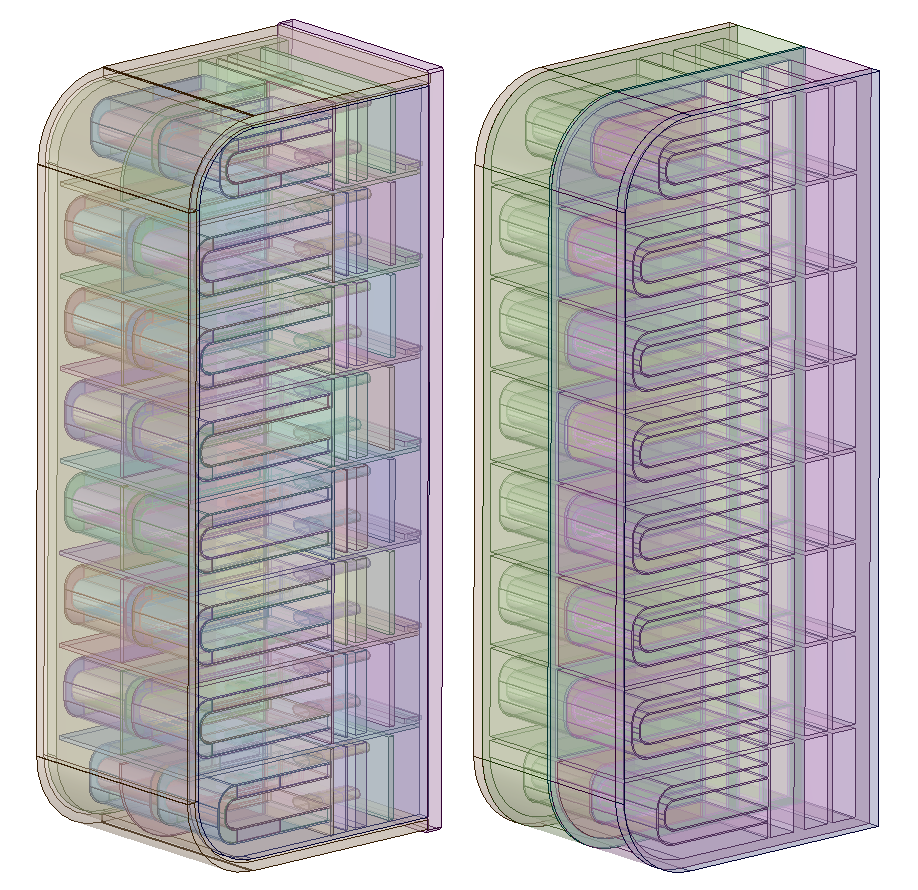}
\hspace{-7.8cm}\raisebox{-0.1cm}{(a)}
\hspace{0.1cm}\raisebox{-0.1cm}{(b)}
\hspace{0.4cm}\raisebox{-0.1cm}{(c)}
\hspace{0.5cm}\raisebox{-0.1cm}{(d)}
\hspace{1.0cm}\raisebox{-0.1cm}{(e)}
\hspace{1.4cm}\raisebox{-0.1cm}{(f)}
\newlength{\lenabcdef}
\settowidth{\lenabcdef}{(a) (b) (c) (d) (e) (f)}
\hspace{4.4cm}\hspace{-1.0\lenabcdef}
\caption{\label{fig:FiComparison} Various 3D models for the ferromagnetic components. (a) a simple FI geometry for fast testing. (b) and (c) are the FI models used for calculations (no difference in results). (d) the original FI configuration model consisting of stacks of 40mm thick plates, which were merged in (c).  \label{fig:TbmComparison} The TBM geometry before (e) and after (f) simplification.}
\end{figure}

We made several simplified FI models starting from the configuration model, that are shown in Fig. \ref{fig:FiComparison}(a-d). The first simplification was to remove radial gaps. The FIs are made of stacks of 40\,mm thick steel plates with 4\,mm gaps between the plates. To expedite the calculations, several toroidal gaps less than 10\,mm wide were removed and a few somewhat wider gaps were extended by 10\,mm. Modifying the gaps did not change the final results. Many small surface details, such as bevels, were also removed to simplify the final model.

\label{sec:TBMgeom}

Three equatorial ports of ITER are dedicated to TBMs, with room for two TBMs in each port. In our study, we placed a model of the European helium-cooled pebble bed (HCPB) TBM into all six TBM slots. In reality the different ITER partners will have their own test blanket module implementations in their slots.

The CAD model for the HCPB is very complex, but we received a model which had cooling ducts and many other smaller features already removed. We further simplified the model by, e.g., removing piping, merging thin plates together and removing air gaps. Figures \ref{fig:TbmComparison}(e-f) show the geometry before and after simplification.

The whole torus, including the coils, was enclosed inside a so called ``finite sphere'' with a radius of 14 meters. This radius allowed us to fit the components inside the sphere with a margin of several meters in most cases, but only about one meter near a PF coil.
The same structure was also implemented in the permanent magnet model albeit with all the coils absent.
The finite sphere was enclosed inside a spherical shell, dubbed the ``infinite shell''. The name comes from the COMSOL feature we used to remap the radial coordinate $\varrho^2=R^2+z^2$ inside the infinite shell in order to use boundary conditions at infinity:
\begin{equation}
\varrho' = \varrho_0\frac{\Delta \varrho}{\varrho_0 +\Delta \varrho - \varrho},\label{eq:inftyMapping}
\end{equation}
where $\varrho_0$ is the inner radius of the infinite shell and $\Delta \varrho$ is the thickness of the shell. In our model the outer surface of the shell was mapped to a distance of several kilometers.

\section{Current densities and material properties for the Finite Element Analysis}

\noindent We included the following free currents in our model: the toroidal and poloidal field coils (including the central solenoid) and the toroidal plasma current.
The magnitude of the current density was assumed to be uniform in each coil.
For the circular PF and CS coils,  the current direction is trivial to calculate, but for the D-shaped TF coil we assumed the direction to be parallel to the spine curve described in section \ref{sec:TFspineCurve}. COMSOL has the functionality to create a curvilinear coordinate system within the TF coil, but as this seemed to cause numerical problems in our geometry, we chose another route: a MATLAB routine returning  a unit vector parallel to the spine curve of the TF coil was created.
\label{sec:plasmaCurrent}
The magnitude of the toroidal plasma current density was calculated with a MATLAB routine from the current flux function $f$ and the pressure flux function $p$ read from the EQDISK file using equation 3.3.6 in \cite{wesson}. (Note: there is a typo in that equation; $\mu_0$ should be in the denominator.)

All domains were assumed to have the same electrical properties: relative permittivity $\varepsilon_r=1$ and conductivity $\sigma=1\,\textrm{S/m}$. The current density was fixed in A/m$^2$. Most domains had linear magnetic response with  relative magnetic permeability $\mu_r=1$, while for the ferromagnetic components we set the magnetization $M$ as a function of magnetizing field $H$ by defining the $H$-$B$ curve, or $H(B)$.

The FIs will be made of SS430 stainless steel, for which the $B$-$H$ curve was the mean curve of the $B$-$H$ curves in the table 4-1 of \cite{HBcurveFI}.  The magnetic properties data for the EUROFER steel \cite{HBcurveTBM} of the TBMs is temperature dependent, but we made a conservative assumption of uniform 350$^\circ$C. A two-piece linear model for the $H$-$B$ curve was constructed from the three available parameters.
The first linear segment was assumed to pass through the origin, and the slope was calculated from the ratio of the coercive field $H_c$ and the remanent magnetization $M_r$. The slope of the high field segment was vacuum permeability $\mu_0$, and the knee point was calculated by solving the location where the first segment passes through  saturation magnetization $M_s$.

Removing small details changed the metal volumes of the TBMs and FIs. To compensate for this, we modified the magnetic response of the materials, i.e., the $\mathbf{H}(|\mathbf{B}|)$ function. We required that the simplifications would not change  the magnetic moment $\mathbf{\mathcal{M}}=\int\mathbf{M}\mathrm{d}V$ of the objects at the uniform magnetic field limit. This resulted in the formula for a new $H$-$B$ curve $H^*(B)$, where the difference in metal volume is accounted for:
\begin{equation}
H^*(B)=H\left(B+\{1-c\}\{B-\mu_0 H(B)\}\right).
\end{equation}
Here the volume ratio $c$ is defined as $c=V_\mathrm{original}/V_\mathrm{simplified}$, and it varies between  $0.7355$ and $0.738$ for the different kinds of FI sectors and is $1.001$ for the TBMs.

\section{Results}

\begin{figure}
\centering
\includegraphics[width=0.47\linewidth,trim=0 0 58mm 0,clip]{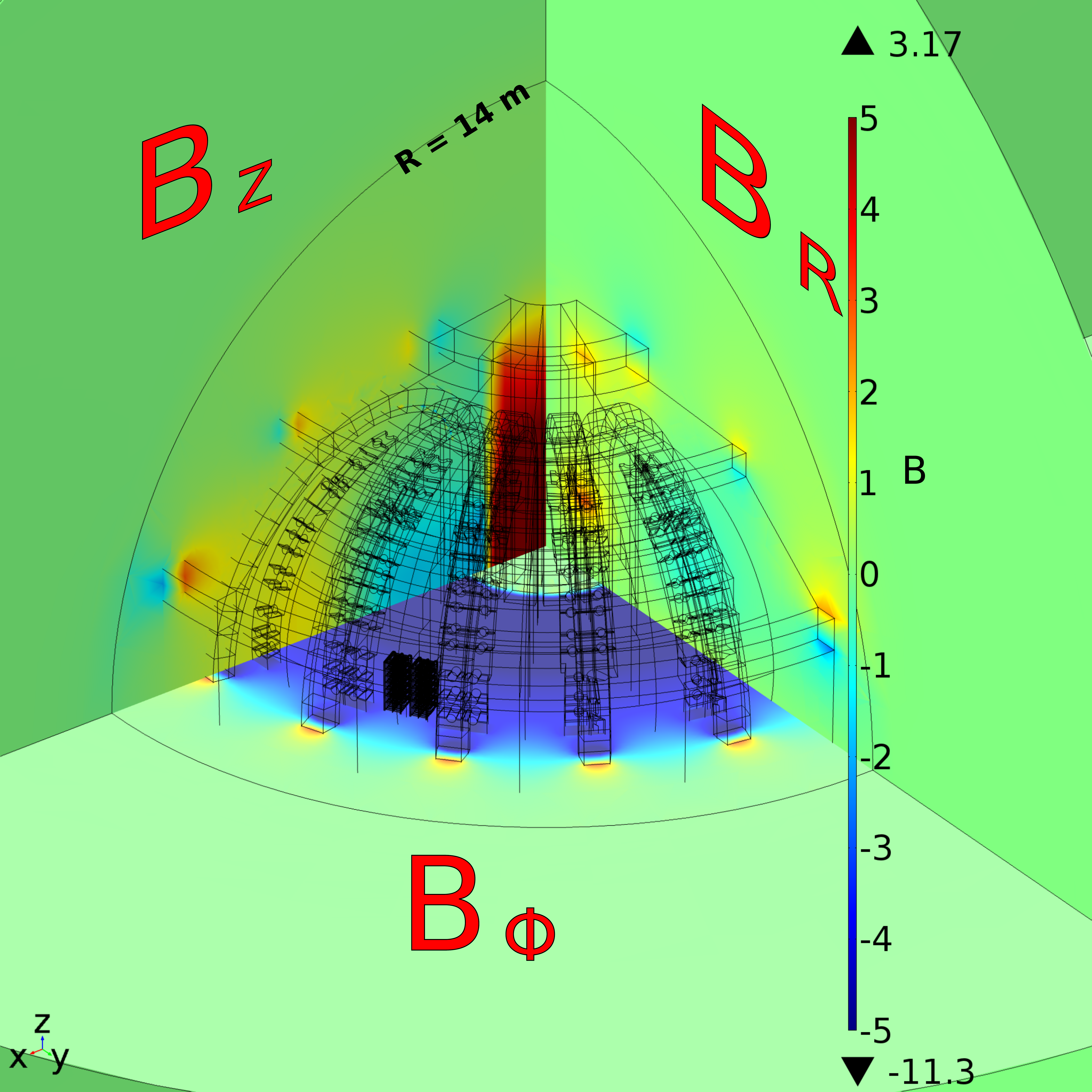}
\includegraphics[width=0.47\linewidth,trim=0 0 58mm 0,clip]{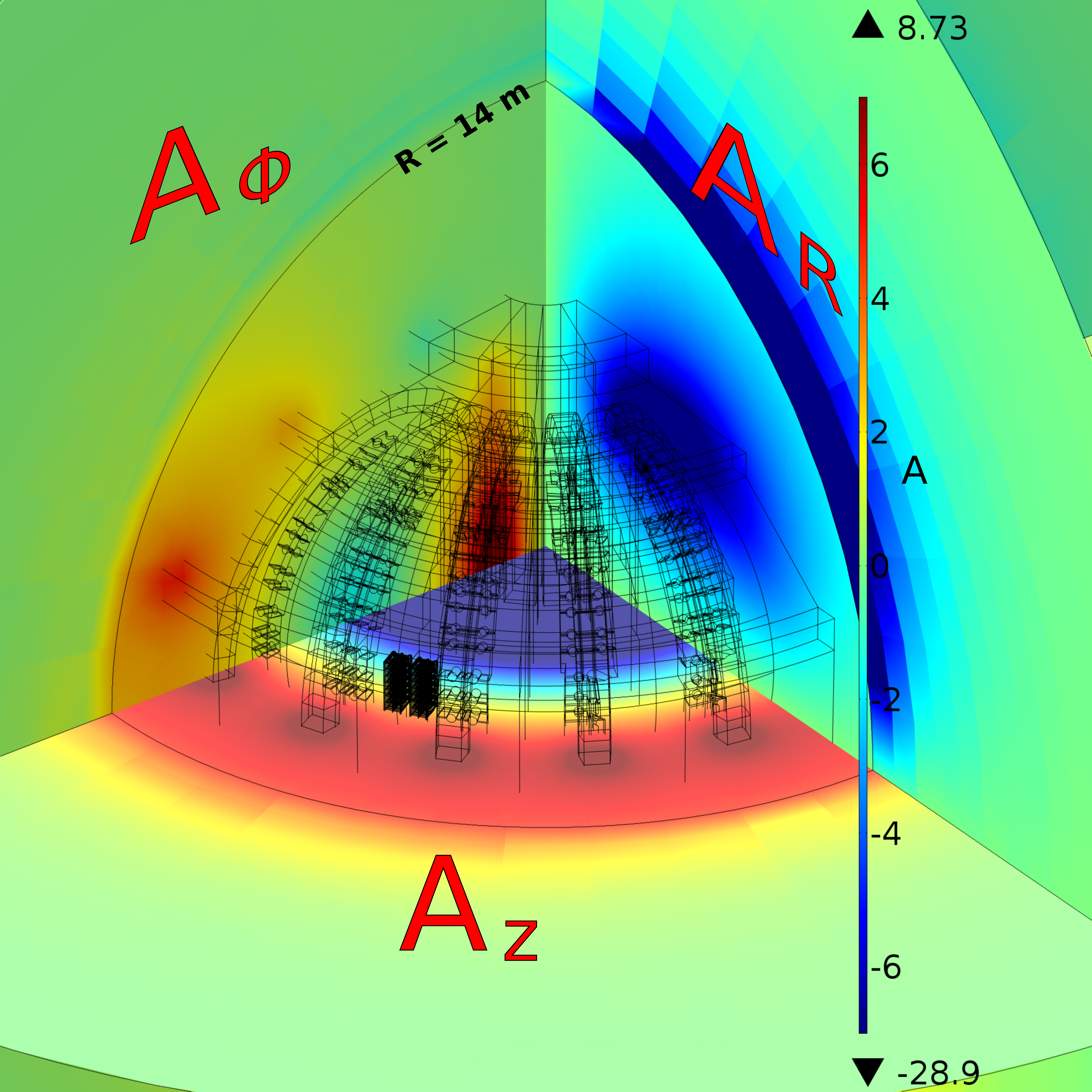}\\
\includegraphics[width=0.47\linewidth,trim=0 0 58mm 0,clip]{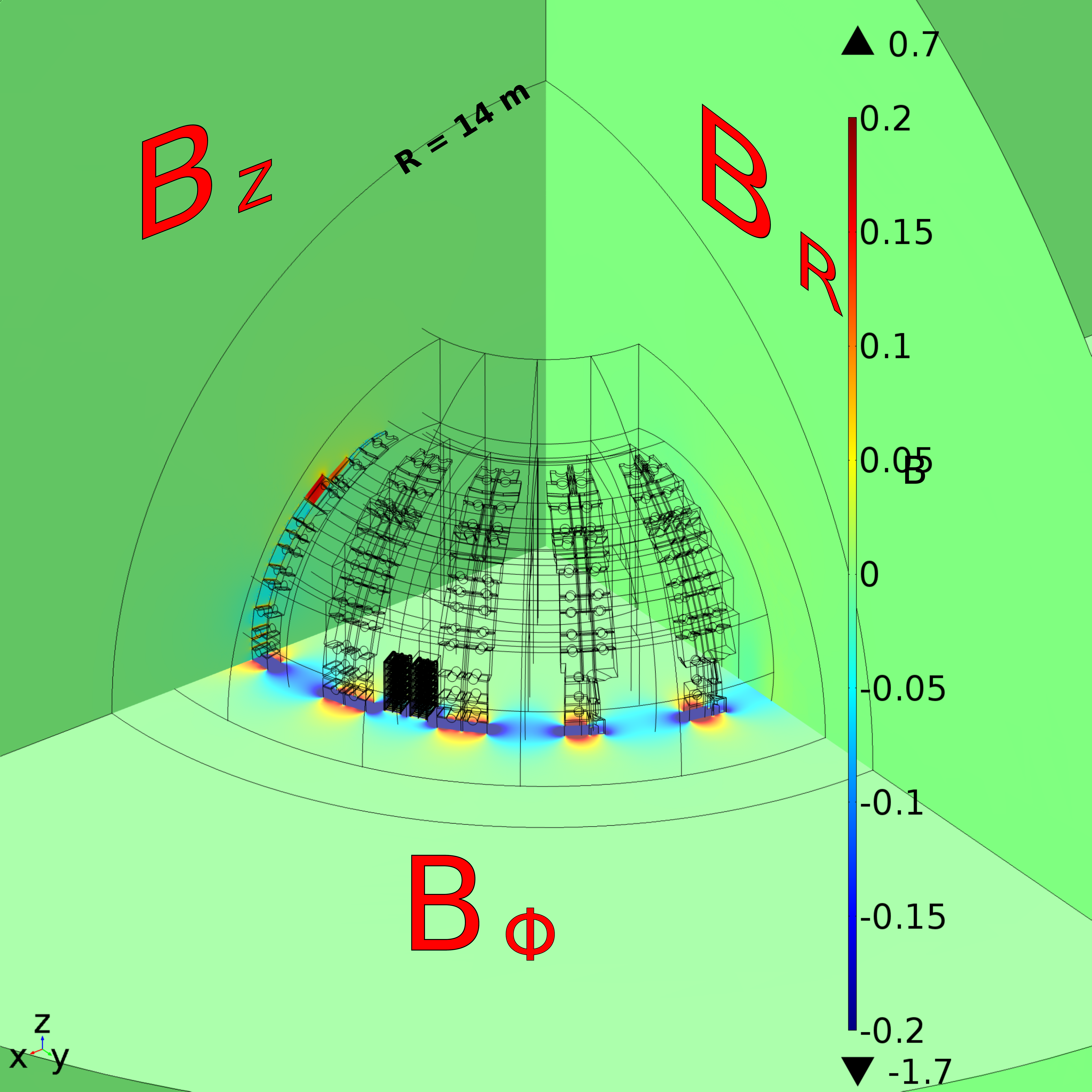}
\includegraphics[width=0.47\linewidth,trim=0 0 58mm 0,clip]{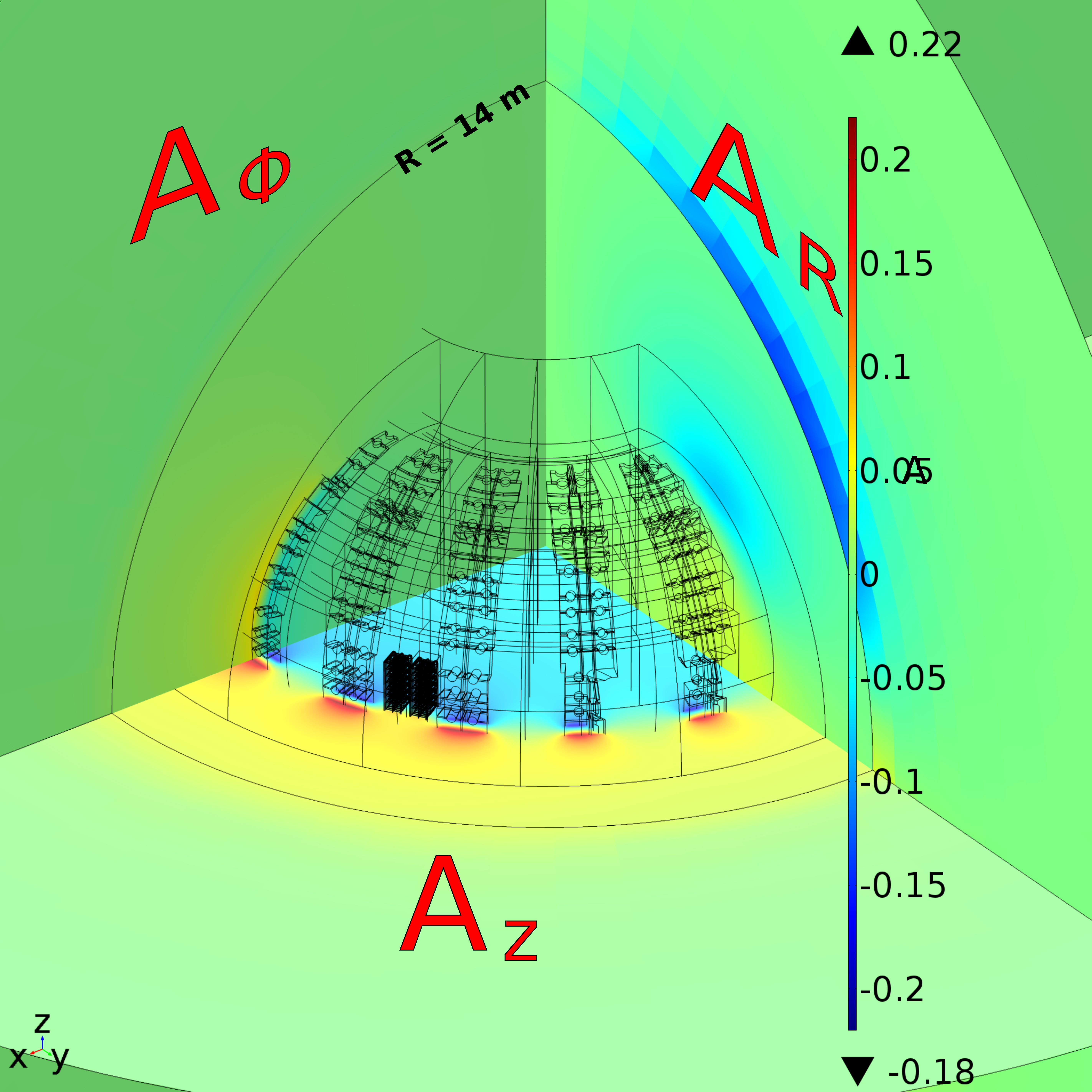}
\caption{\label{fig:cutplanes}The magnetic flux density $\mathbf{B}$ (T) and the total magnetic vector potential $\mathbf{A}$ (T/m)  as calculated by COMSOL. The total field is shown on the top row and the field due to the ferromagnetic components on the lower row.  All images show three orthogonal cut planes displaying three orthogonal components of the field, as indicated in the figure. The thin black lines indicate component and domain boundaries in the model. There are harmless numerical artifacts visible in $A$ in the ``infinite shell'', caused by remapping of the spatial coordinates.}
\end{figure}

\begin{figure}
\centering
\raisebox{3.6cm}{(a)}\includegraphics[height=0.5\linewidth]{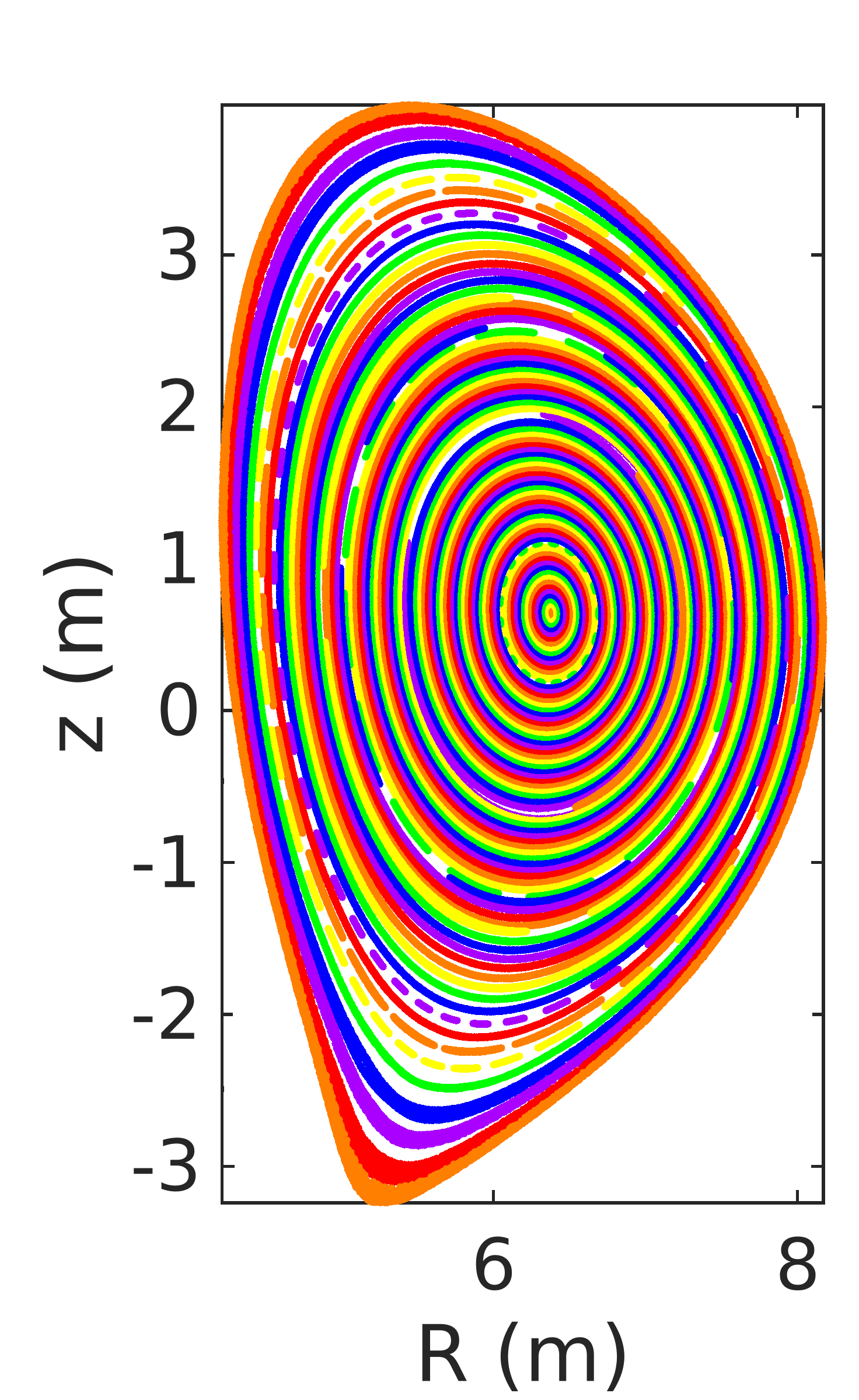}%
\raisebox{3.6cm}{(b)}\includegraphics[height=0.5\linewidth]{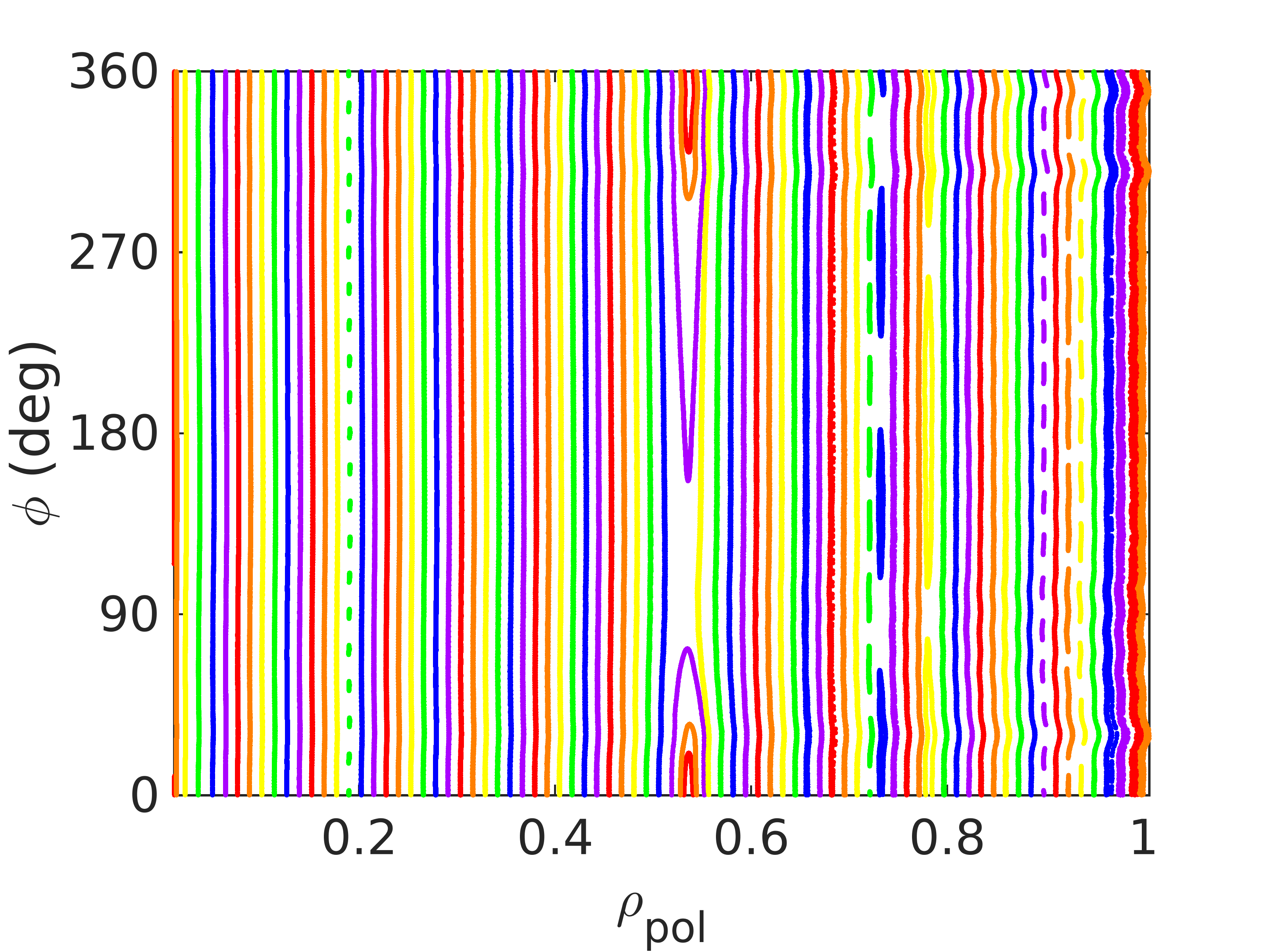}\\%
\raisebox{3.6cm}{(c)}\includegraphics[height=0.5\linewidth]{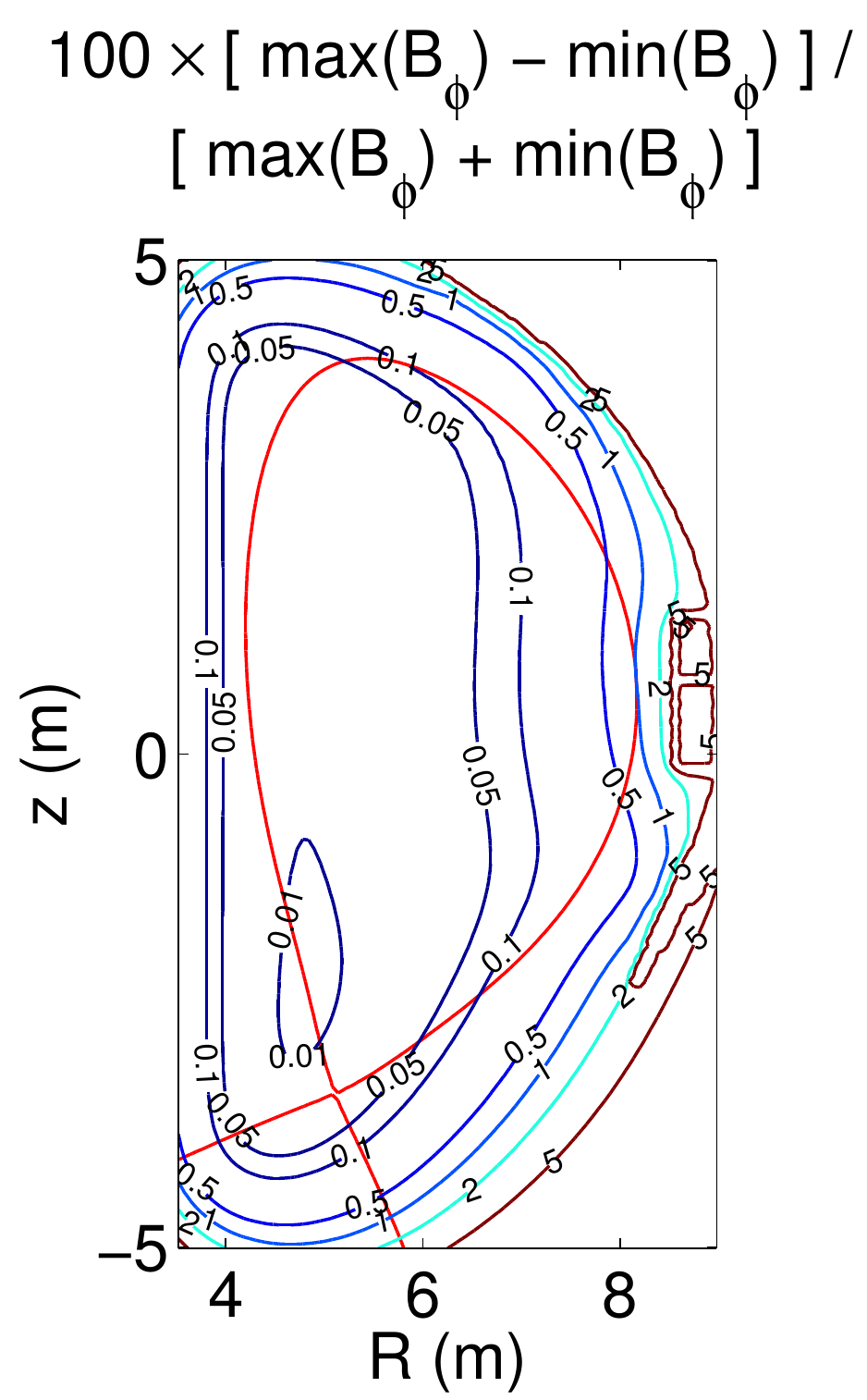}%
\raisebox{3.6cm}{(d)}\includegraphics[height=0.5\linewidth]{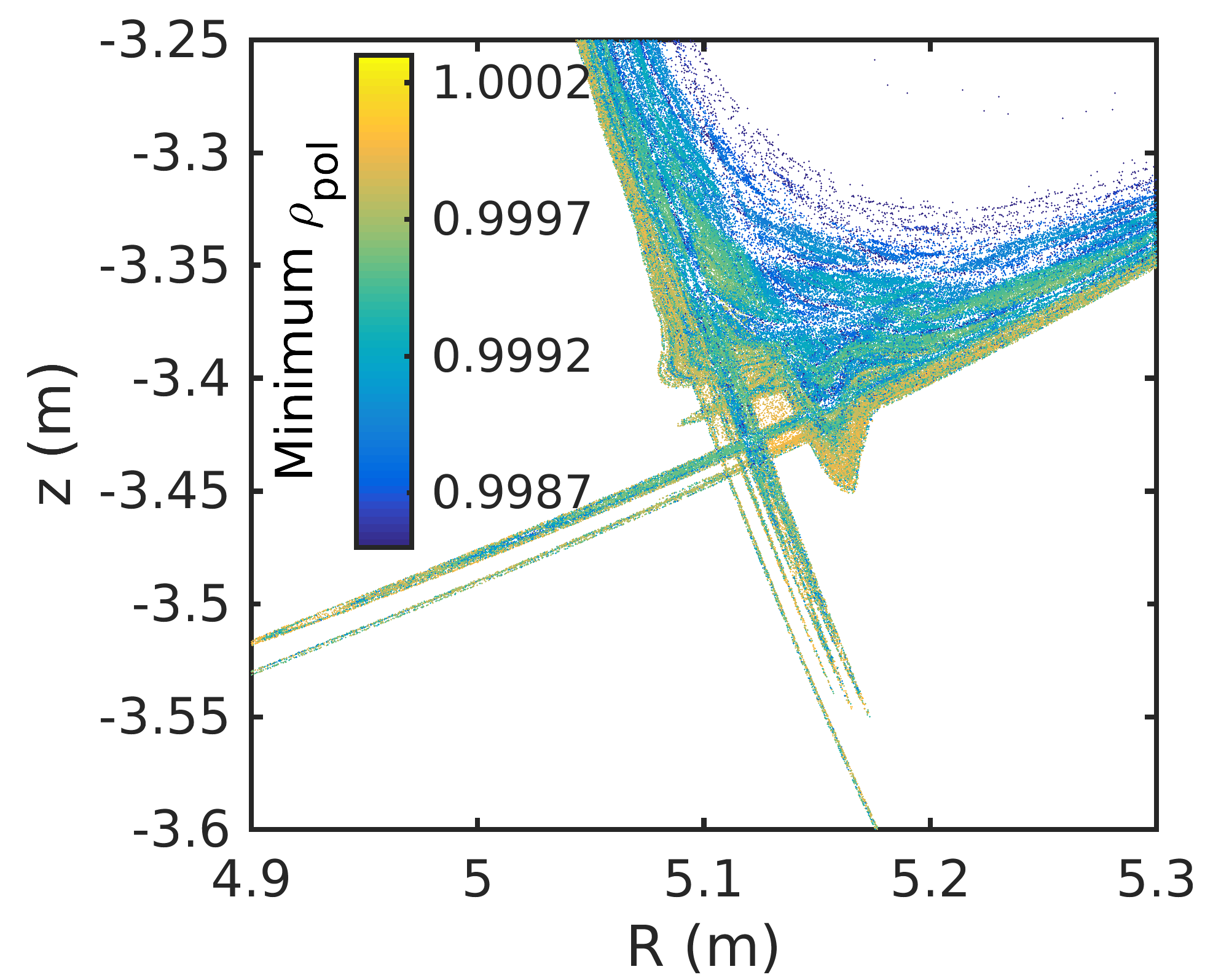}
\caption{\label{fig:ripplemaps} (a) Poloidal and (b) toroidal Poincaré plot showing the induced island structure within the plasma, (c) toroidal field ripple map, (d) the homoclinic tangle near X-point. }
\end{figure}

\begin{figure}
\centering%
\raisebox{2.3cm}{(a)}\hspace{-1.5em}\includegraphics[height=0.34\linewidth]{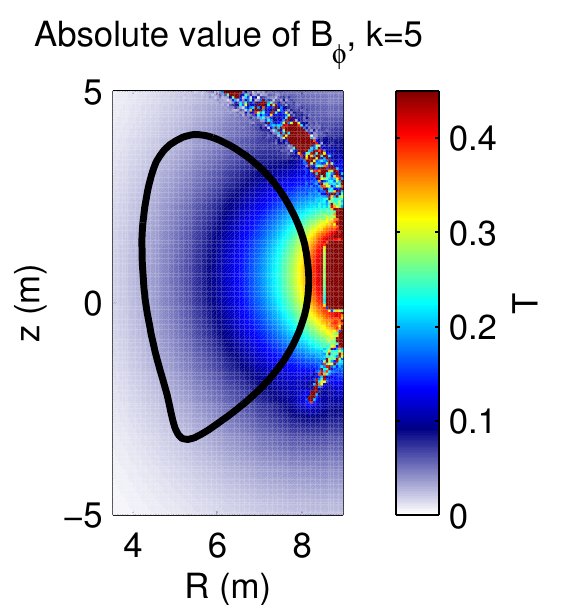}
\raisebox{2.3cm}{(b)}\hspace{-1.5em}\includegraphics[height=0.34\linewidth]{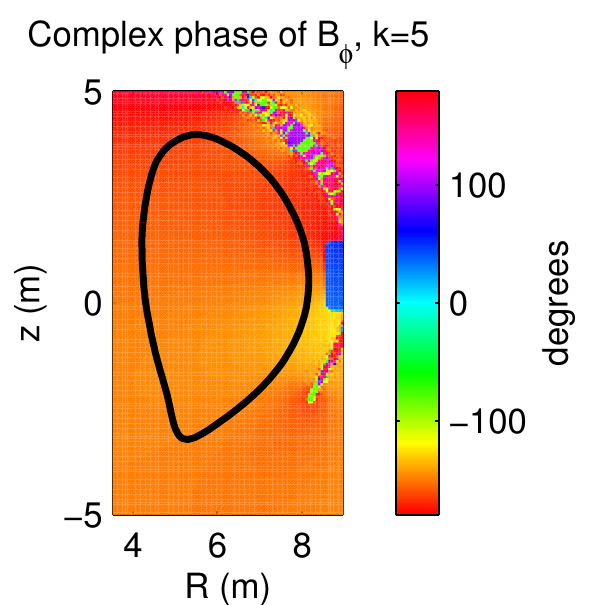}
\includegraphics[height=0.34\linewidth]{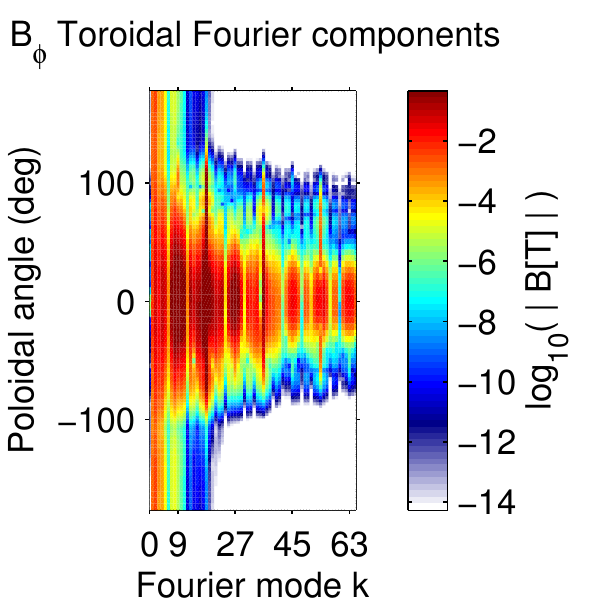}\hspace{-4.8em}\raisebox{2.3cm}{(c)}\\
\raisebox{2.3cm}{(d)}\hspace{-1.5em}\includegraphics[height=0.34\linewidth]{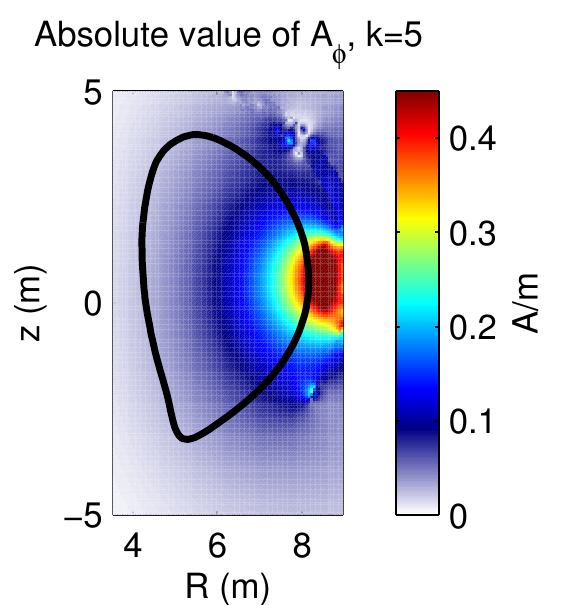}
\raisebox{2.3cm}{(e)}\hspace{-1.5em}\includegraphics[height=0.34\linewidth]{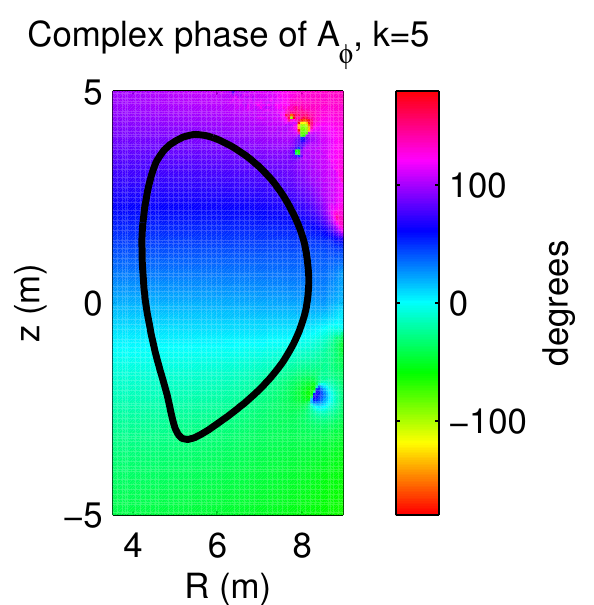}
\includegraphics[height=0.34\linewidth]{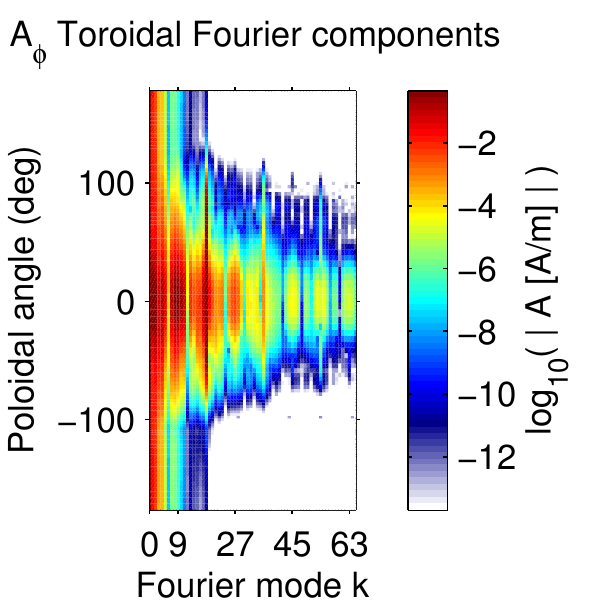}\hspace{-4.8em}\raisebox{2.3cm}{(f)}%
\caption{\label{fig:fft} Toroidal Fourier decomposition of the toroidal components of $\mathbf{B}$ and $\mathbf{A}$ fields. A single component on a poloidal plane is shown in figures (a), (b), (d) and (e). Figures (c) and (f) show the amplitude of the first 65 modes along the separatrix  (black line).  }
\end{figure}

\noindent The COMSOL calculation produced 3D magnetic flux density $\mathbf{B}$ and vector potential $\mathbf{A}$ for various ITER scenarios. Figure \ref{fig:cutplanes} shows these fields for the 15\,MA plasma current H-mode scenario during flat top phase. We then combined the COMSOL results with Biot-Savart law integrated background fields and analysed the field. Figure \ref{fig:ripplemaps} shows a toroidal field ripple map (a measure of FI performance), the homoclinic tangle near the X-point, and Poincaré plots showing the structure inside the plasma separatrix. Finally, the field was decomposed into toroidal Fourier components, which are shown in Figure \ref{fig:fft}. These form the input for spectral plasma response codes. The magnetic fields were verified against earlier work \cite{ITER_D_64D8GV} and were found to agree quantitatively. Please check the supplementary material (available electronically) for illustration.

\section{Summary and Future work}
\noindent
 We have devised a method for calculating the magnetization of ITER ferromagnetic components using the finite element method, and combined the resulting magnetic flux density $\mathbf{B}$ and magnetic vector potential $\mathbf{A}$ to Biot-Savart law integrated fields. The former can be used for studying, e.g., fast ion behaviour in the realistic ITER 3D field, while the latter provides the input for calculating the plasma response to the external perturbation.

\paragraph{Future work}  We are using the calculated fields in the ASCOT~\cite{ascot4ref} suite of codes in order to simulate fast ion wall loads due to fusion alphas and heating neutral beams.

The simulations are in progress, and we already reported that the detail level at which the FIs are modeled in the field calculation does not appear to play a significant role for fusion alpha wall loads at least in the 15\,MA H-mode case \cite{tainaEps2014,tainaIaeaFec2014}. The 3D fields will also be delivered to our collaborators so that the plasma response to the external perturbations can be included in future wall load simulations. At the time of writing this article, we have not discovered strong changes in the wall loads due the European TBMs. Only a couple of ITER scenarios have so far been addressed. %

\section*{Acknowledgements}
\noindent The authors would like to thank Nicolò Marconato and Pablo Vallejos for their help on using COMSOL.

{%
This project has received funding from Fusion For Energy (grant F4E-GRT-379), the Academy of Finland (project No. 259675), Tekes – Finnish Funding Agency for Technology and Innovation.
We acknowledge the computational resources from Aalto Science-IT project, CSC - IT center for science and the International Fusion Energy Research Centre.
}

\section*{References}
\bibliography{bibfile}

%
%
%
\end{document}